\documentclass[11pt,a4paper]{article}
\usepackage{jheppub}
\usepackage{amsmath}
\usepackage{amssymb}
\usepackage{tikz}
\usepackage{pgfplots}
\usepackage{graphicx}

\newcommand{\M}{\mathcal{M}}
\newcommand{\q}{\tilde{q}}
\newcommand{\E}{\mathcal{E}}

\newcommand{\nGamma}{\widehat{\Gamma}}

\newcommand{\be}{\begin{equation}}
\newcommand{\ee}{\end{equation}}
\newcommand{\bea}{\begin{eqnarray}}
\newcommand{\eea}{\end{eqnarray}}

\def\({\left(} \def\){\right)}
\def\[{\left[} \def\]{\right]}

\title{Non-relativistic geometry of holographic screens}
\author{Mudassir Moosa}

\affiliation{Center for Theoretical Physics and Department of Physics,\\
University of California, Berkeley, CA 94720, U.S.A.} 
\affiliation{Lawrence Berkeley National Laboratory, Berkeley, CA 94720, U.S.A.} 

\emailAdd{mudassir.moosa@berkeley.edu}

\abstract{We propose that the intrinsic geometry of holographic screens should be described by the Newton-Cartan geometry. As a test of this proposal, we show that the evolution equations of the screen can be written in a covariant form in terms of a stress tensor, an energy current, and a momentum one-form. We derive the expressions for the stress tensor, energy density, and momentum one-form using Brown-York action formalism.

}

\begin{document}
	\maketitle
	\section{Introduction} \label{sec-intro}
	%
	The holographic principle \cite{hooft-hp,susskind-hp}, loosely speaking, asserts that the physical degrees of freedom of quantum gravity in a region of area $A$ can not be more that $\frac{A}{4G\hbar}$. This principle was presented in a more formal setting in terms of covariant entropy bound, or Bousso bound \cite{Bousso-1999,Flanagan:1999jp,Bousso-2002,Bousso:2014sda,Bousso:2014uxa}, which bounds the entropy on the lightsheets of any spatial codimension-$2$ surface by the area of this surface. The AdS-CFT correspondence \cite{ads-cft} provides an example of this principle. The correspondence states that the quantum state on a slice of AdS is described by the state of the CFT on the boundary of that slice.
	
	Despite the success of the AdS-CFT correspondence, not much progress has been made in generalizing it to general spacetimes. The reason behind the lack of progress in this direction is that it is not clear where the information of the state of the bulk theory on a given slice of the spacetime lives. That is, it is not clear what codimenion-$1$ surface plays the role of the conformal boundary of the AdS. Holographic screens  \cite{Bousso-99} are believed to be a candidate for this surface.
	
	A holographic screens is constructed by starting with a null foliation, $N(R)$, of the spacetime. On each null hypersurface, we choose a spatial cross-section of largest area. This cross-section, $\sigma(R)$, called \emph{leaf}, is a codimension-$2$ spatial surface with vanishing expansion. That is,
	\begin{equation}
	\theta^{(k)} = \, 0 \, ,
	\end{equation}
	where $k^{a}$ is the null generator of the null slice, $N(R)$. 
	Assuming the null energy condition, the focussing theorem (see for \emph{e.g.} \cite{Wald}) guarantees that the portions of $N(R)$ `outside' and `inside' of $\sigma(R)$ are two seperate lightsheets of $\sigma(R)$. The covariant entropy bound then implies that the entropy of the quantum state on the null slice is bounded by the area of the corresponding leaf. The codimension-$1$ hypersuface foliated by these leaves, by construction, bounds the entropy of the spacetime, slice by slice. This codimenion-$1$ hypersurface is called holographic screen.

	Recently it has been shown that the area of the leaves obey an area law in the spacetimes which satisfy few generic conditions \cite{bousso-engelhardt-small,bousso-engenhardt}. The area law states that the area of the leaves is monotonic as we translate from one leaf to another. The fact that the area of leaves is not constant seems problematic, as it indicates that the degrees of freedom of the dual theory vary with time. This suggests that the dual theory might not be a unitary QFT.
	
	
	Some progress has been made in developing the holographic theory of the general spacetimes. In \cite{weinberg-sanches}, the holographic entropy formula of \cite{RT,HRT} has been generalized beyond AdS-CFT correspondence. They proposed that the entanglement entropy of the reduced quantum state on some subregion of the leaf is related to the area of the stationary surface anchored on the boundary of that subregion. This proposal has been used in \cite{nomura-1,nomura-2} to study the Hilbert space structure of the holographic theory.
	
	In \cite{bousso-moosa}, we have studied the dynamical equations of holographic screen. These are actually the constraints equations of GR (see Eqs.~\eqref{eq-einstein}), 
	but we have decomposed them in $2+1$ equations, using the additional structure of the screen in terms of foliation into leaves. We will review these equations in  Sec.~(\ref{sec-screen-rev}). These $2+1$ equations are invariant under the foliation preserving diffeomorphism, but not under arbitrary coordinate transformation on the screen. One of the main results of this paper is to write a covariant version of these equations. 
	
	The goal of this project is to understand the intrinsic geometry of holographic screen, $H$. The motivation behind this is that if we hope to put a holographic field theory on $H$, then we need to understand how to couple this theory with the geometry of $H$. For instance, we have to define the notion of parallel transport and covariant derivative on $H$. 
	%
	If $H$ were a hypersurface which is nowhere null, with the normal vector normalized such that $\hat{n}^{a}\hat{n}_{a} = \pm 1$, then the induced metric on the screen would have been
	\begin{equation}
		\gamma_{ab} = g_{ab} \mp \hat{n}_{a}\hat{n}_{b} \, . \label{eq-hs-metric}
	\end{equation}
	In this paper, we will assume the following index notation: The first half of the Latin letters $(a,b,c)$ denote the directions in the full $4$-dimensional spacetime. The second half of the Latin letters $(i,j,k)$ denote the directions on the codimension-$1$ hypersurface (such as $H$). The upper case Latin letters $(A,B,C)$ denote the directions on the codimenison-$2$ surface (such as $\sigma(R)$). Let $e_{i}^{a}$ be the pull-back operator that pull backs an one-form from the spacetime to the codimension-$1$ hypersurface. Similarly, we define $e_{A}^{a}$ and $e_{A}^{i}$. The pull-back of the spacetime metric on $H$ is
	\begin{align}
		\gamma_{ij} =& \, g_{ab}e^{a}_{i}e^{b}_{j} \, ,\\
		=& \, \gamma_{ab}e^{a}_{i}e^{b}_{j} \, .
	\end{align} 	
	This metric would have completely described the intrinsic geometry of $H$. For instance, if we were only given $H$ with this metric, and had no information of the geometry of  background spacetime, we could have used this metric to define various geometric quantities on $H$, like connection and intrinsic Riemann curvature for example. 
	
	
	However, there are known examples where $H$ does not have a fixed signature. As a result, the induced metric becomes degenerate at points where the hypersurface changes signature. Therefore, we can not invert this metric, which means we can not use the standard formula of various geometric quantities, such as Christoffel symbols.\footnote{Of course, one can still define various quantities as the pull back of their analogues from the full spacetime. However, this would require the knowledge of the background spacetime, where as we want to study the geometry of $H$ assuming we know nothing about the metric of the spacetime.} This concludes that the metric in Eq.~\eqref{eq-hs-metric} is not suitable to describe the intrinsic geometry of $H$.
	
	Furthermore, holographic screens are not arbitrary hypersurfaces of indefinite signature. Rather they have an additional structure of foliation into leaves. It was shown in \cite{bousso-engenhardt} that this foliation is unique. This hints to the non-relativistic structure of holographic screen. Another reason to expect the non-relativistic structure of holographic screen comes from the fact that the quantization on the light front is non-relativistic (see for \emph{e.g.} \cite{Bigatti:1997jy})\footnote{We thank R. Bousso for emphasizing these points.}. 
		
	Motivated by these observations, we propose that the geometry of holographic screen should be described by a non-relativistic geometry. In Sec.~(\ref{sec-nc-rev}), we will review the Newton-Cartan (NC) geometry, an example of a non-relativistic geometry. In Sec.~(\ref{sec-nc-on-screen}), we will argue that the Riemannian metric on the leaves, the foliation structure of the screen, and the existence of the vector field normal to the leaves but tangent to the screen, allow us to define NC data on $H$, which we can use to define connection and curvature on $H$. 
	
	In Sec.~(\ref{sec-screen-eqs}), we will use the diffeomorphism invariance of the screen to derive a covariant version of the dynamical equations of the holographic screens. This will be a test of our proposal that the geometry of the screen is described by the NC formalism. These covariant equations will be in terms of a screen `stress tensor', `energy current', and `momentum one-form'. We will make an ansatz of these tensors in Sec.~(\ref{sec-screen-eqs}). Finally, we will follow \cite{booth-fairhurst} to perform the treatment similar to that of Brown and York \cite{brown-york} to verify our ansatz.
	
	
	\section{Holographic screens} \label{sec-screen-rev}
	
	
	A \emph{future/past} holographic screen \cite{bousso-engelhardt-small,bousso-engenhardt} is constructed by starting with the world line of an observer and  shooting future/past light cones from it \cite{Bousso-99}. These light cones provide us with an one-parameter null foliation, $N(r)$, of some region of the spacetime. On each null hypersurface, we choose the spatial cross-section of maximum area. These spatial surfaces are called leaves, $\sigma(r)$. The Bousso bound \cite{Bousso-1999,Bousso-2002} guarantees that entropy on $N(r)$ is bounded by the area of $\sigma(r)$. The union of these leaves is called the future/past holographic screen. 
	
	By construction, holographic screen is a codimension-$1$ hypersurface, which is foliated by the codimension-$2$ spacelike surfaces, $\sigma(r)$, where $r$ is the foliation parameter. Since the background manifold is Lorentzian, there are two future directed null vectors normal to the leaves, $l^{a}$ and $k^{a}$. We fix their relative normalization by demanding $k^{a}l_{a}=-1$. The defining property of holographic screen is that the expansion of the null congruence in $k^{a}$ direction vanishes, that is
	\begin{equation}
		\theta^{(k)} \equiv \tilde{q}^{ab}\nabla_{a}k_{b} = 0 \, , \label{theta-k}
	\end{equation}
	where 
	\begin{align}
		\tilde{q}^{ab} =& \, g^{ab}+k^{a}l^{b}+l^{a}k^{b} \, ,\\
		=& \, e^{a}_{A}e^{b}_{B}\tilde{q}^{AB} \, ,
	\end{align}
	is the induced (inverse) metric on the leaves. For future/past holographic screens, the leaves are marginally trapped/anti-trapped \cite{bousso-engelhardt-small,bousso-engenhardt}. This means that the expansion in the $l^{a}$ direction is negative for future holographic screens,
	\begin{equation}
	\theta^{(l)} < 0 \, ,
	\end{equation}
	where as the opposite inequality holds for past holographic screen.
	
	Given an arbitrary codimension-$1$ hypersurface of indefinite signature, one might ask what data is required to describe the intrinsic geometry of the hypersurface. One candidate is to use the induced metric on the hypersurface, defined as the pull-back of the spacetime metric
	\begin{equation}
		\gamma_{ij} = \, e_{i}^{a}e_{j}^{b} g_{ab} \, . \label{eq-deg-met}
	\end{equation}
	Though this works for the hypersurfaces that are nowhere null, it is not suitable for the hypersurfaces which do not have a fixed signature. For these hypersurfaces, the metric in Eq.~\eqref{eq-deg-met} becomes degenerate, and hence non-invertible, wherever the signature changes. As a result, we can not define connection, curvature, or any other intrinsic quantity that requires the inverse metric, on these hypersurfaces. 
	
	As holographic screens do not have a definite signature, it is interesting to understand how to describe their intrinsic geometry. In this section, we will review some properties of the holographic screens that were proven in \cite{bousso-engenhardt}. We will see in Sec.~(\ref{sec-nc-on-screen}) that these properties allow us to propose that the intrinsic geometry of the holographic screens can be described by the Newton-Cartan geometry. We will also review the evolution equations of the holographic screens, that were studied in \cite{bousso-moosa}, in the current section.
	
	Let $h^{i}$ be the vector tangent to the screen, but normal to the leaves. This means that we can write it as a linear combination \cite{bousso-engelhardt-small,bousso-engenhardt}
	\begin{equation}
		e_{i}^{a}h^{i} = h^{a} = \alpha l^{a} + \beta k^{a} \, . \label{eq-lc-h}
	\end{equation}
	Since the signature of the screen is not fixed, we can not normalize the vector $h^{a}$ by fixing its norm. Instead, we normalize it by demanding that $h^{a}(dr)_{a}= h^{i}(dr)_{i} = 1$. Similarly, we fix the length of normal vector, $n^{a}$, to the holographic screen by demanding $n^{a}n_{a} = - h^{a}h_{a}$. This implies that the normal vector can be written as
	\begin{equation}
		n^{a} = -\alpha l^{a} + \beta k^{a} \, .
	\end{equation}
	
	 
	Assuming some mild generic conditions, an area law for the holographic screens was proven in \cite{bousso-engelhardt-small,bousso-engenhardt}. An important ingredient in the proof of this area law in \cite{bousso-engenhardt} was the proof of the fact that $\alpha$ in Eq.~(\ref{eq-lc-h}) does not vanish. In particular, it was shown that $\alpha < 0$ for future holographic screens, while $\alpha>0$ for past screens. This important result, together with Eq.~\eqref{theta-k} and Eq.~\eqref{eq-lc-h}, immediately yields the area law: ``the area of the leaves always increases as we move along the screen''. That is,
	\begin{equation}
	\theta^{(h)} = \alpha \theta^{(l)} > 0 \, .
	\end{equation}
		The result $\alpha \ne 0$ is important to us for one another reason. We will see in Sec.~(\ref{sec-nc-on-screen}) that this result plays a significant role in the consistency of our proposal to describe the intrinsic geometry of the screen using Newton-Cartan geometry. 
	
	
	
	Another important theorem proven in \cite{bousso-engenhardt} guarantees the uniqueness of the foliation of a holographic screen into leaves. In particular, it was shown that if we take a screen and choose a different foliation into codimension-$2$ closed surfaces, these surfaces will not have vanishing null expansion.	
	This implies that the foliation into leaves is important for the defining property of the holographic screens, as was also emphasized in \cite{booth-fairhurst}. Without this foliation, $\theta^{(k)} = 0$ condition is not well-defined. 
		
	
	Imposing Einstein equations on the spacetime puts constraints on the extrinsic geometry of a codimesion-$1$ hypersurface,
	\begin{equation}
	G_{ab}n^{a}e^{b}_{i} = 8\pi T_{ab}n^{a}e^{b}_{i} \, . \label{eq-einstein}
	\end{equation}
	For hypersurfaces with fixed signature, Gauss-Codazzi equations (see for \emph{e.g.} \cite{toolkit}) can be used to write these constraint equations as
	\begin{equation}
		\widehat{D}_{j}(\widehat{K}_{i}^{\, j} - \gamma_{i}^{\, j}\widehat{K}) = T_{ab}\hat{n}^{a}e_{i}^{b} \, , \label{rel-open-sys}
	\end{equation}
	where $\widehat{D}_{j}$ is the covariant derivative compatible with the metric $\gamma_{ij}$ in Eq.~\eqref{eq-deg-met}, and $\widehat{K}_{ij}$ is the extrinsic curvature of the hypersurface. 
		However, Eq.~\eqref{rel-open-sys} is not well-defined for the holographic screens as both the covariant derivative compatible with $\gamma_{ij}$ and the extrinsic curvature tensor are defined using the inverse induced metric, $\gamma^{ij}$. To resolve this problem, it was suggested in \cite{bousso-moosa} to use the foliation structure of the screen, and to use the geometric quantities defined using the (Riemannian) metric on the leaves, such as the extrinsic curvature of the leaves
	\begin{align}
	B^{(h)}_{AB} =& \, e_{A}^{a}e_{B}^{a} B^{(h)}_{ab} \, ,\\
	B^{(n)}_{AB} =& \, e_{A}^{a}e_{B}^{a} B^{(n)}_{ab} \, ,
	\end{align}
	with 
	\begin{align}
	B^{(h)}_{ab} =& \frac{1}{2} \, \tilde{q}_{a}^{\, c}\tilde{q}_{b}^{\, d}\mathcal{L}_{h}\tilde{q}_{cd} \, ,\\
	B^{(n)}_{ab} =& \frac{1}{2} \, \tilde{q}_{a}^{\, c}\tilde{q}_{b}^{\, d}\mathcal{L}_{n}\tilde{q}_{cd} \, .
	\end{align}
	
	The extrinsic geometry of the screen is not completely fixed by that of the leaves, as $B^{(n)}_{AB}$ has only three components while there are six components of $\widehat{K}_{ij}$. The rest of the information of the extrinsic geometry of the screen is given in terms of the \emph{normal one-form}, 
	\begin{equation}
	\omega_{i} \equiv - e_{i}^{a} l^{b}\nabla_{a}k_{b} \, ,
	\end{equation} 
	which can be decomposed into a scalar and a one-form on the leaf,
\begin{align}
\tilde{\kappa} =& \, h^{i}\omega_{i} = -l^{b}h^{a}\nabla_{a}k_{b} \, ,\\
\Omega_{A} =& \, e_{A}^{i} \omega_{i} = - e_{A}^{a}l^{b}\nabla_{a}k_{b} \, . \label{omega}
\end{align}
	
	In terms of these geometric quantities, and using the foliation structure of the screen, Eq.~\eqref{eq-einstein} can be decomposed in $2+1$ `evolution' equations of the holographic screens \cite{bousso-moosa} (also see \cite{Gourgoulhon:2005ch,Gourgoulhon:2006uc,Gourgoulhon:2008pu} for similar equations for dynamical or trapping horizons \cite{Hayward:1993wb,Ashtekar:2002ag,Ashtekar:2003hk,Ashtekar:2004cn})
	\begin{align}
	\alpha \mathcal{L}_{h}\theta^{(l)} + \theta^{(h)2} - B^{(h)AB}\left(B^{(n)}_{AB} - \theta^{(n)}\tilde{q}_{AB} - \tilde{\kappa}\tilde{q}_{AB}\right) + \widetilde{D}_{A}\widehat{\Omega}^{A} =& \, 8\pi T_{ab}n^{b}h^{a} \, ,\label{screen-eq-scalar}\\
	\tilde{q}_{A}^{\, B}\mathcal{L}_{h}\Omega_{B} + \theta^{(h)}\Omega_{A} + \widetilde{D}_{B}\left( B_{A}^{(n)B} - \theta^{(n)}\tilde{q}_{A}^{\, B} - \tilde{\kappa}\tilde{q}_{A}^{\, B}  \right) - \theta^{(l)}\widetilde{D}_{A}\alpha =& \, 8\pi T_{ab}n^{b}e^{a}_{A} \, , \label{screen-eq-vector}
	\end{align}
	where $\widetilde{D}_{A}$ is the covariant derivative compatible with $\tilde{q}_{AB}$, and $\widehat{\Omega}_{A} \equiv \, e_{A}^{a}\widehat{\Omega}_{a}$ is defined as,
	\begin{align}
		\widehat{\Omega}_{a} \equiv& \, \,  h^{b} \q_{a}^{\, c}\nabla_{c}n_{b} \, .
	\end{align}
	
	These evolution equations are invariant under diffeomorphism on the screen that preserve the foliation structure of the screen. These diffeomorphism include the diffeomorphism on the leaves, plus the reparameterization of the foliation parameter \cite{bousso-moosa}. In Sec.~(\ref{sec-screen-eqs}), we will use the Newton-Cartan structure on the screen to write these equations in a covariant form, that is in a form which is invariant under the full diffeomorphism on the screen.

	\section{Newton-Cartan geometry} \label{sec-nc-rev}
	
	The Newton-Cartan geometry is an example of a non-relativistic geometry, originally developed by Cartan as the geometric formulation of Newtonian gravity \cite{Cartan1923,Cartan1925}. This has gained popularity in the recent years owing to its many applications. For instance, it has been used to write the effective field theory  and to study the covariant Ward identities of the quantum Hall effect in \cite{son-13,son-14}. It has also been used to couple non-relativistic field theories with the background spacetime in \cite{Banerjee:2014nja,Jensen:2014aia,Hartong:2014pma,Geracie:2015xfa,Banerjee:2016laq}, and to present the covariant formaulation of non-relativistic hydrodynamics in \cite{Jensen:2014ama,Geracie:2015xfa,Mitra:2015twa}. It was shown in \cite{Hartong:2015zia} that making the Newton-Cartan geometry dynamical gives rise to Ho\v{r}ava-Lifshitz gravity \cite{horava}, and it has been used to study Lifshitz holography in \cite{Christensen:2013lma,Christensen:2013rfa,Hartong:2014oma,Hartong:2014pma}.
	
	In this section, we present a review of the Newton-Cartan (NC) geometry. There are different conventions used in the literature when describing NC geometry. We will be using the conventions used in~\cite{son-13}. Consider a $2+1$-dimensional manifold $\M$ such that $\M = R \times \Sigma$, where $\Sigma$ are $2$-dimensional Riemannian manifolds (generalization to higher dimensions is trivial). The NC structure on $\M$ means that there is a one-form field $\tau_{i}$ such that 
	\begin{equation}
	\tau \wedge d\tau = 0 \, , \label{eq-cond-tau}
	\end{equation}
	a degenerate inverse metric $h^{ij}$ such that
	\begin{equation}
		h^{ij}\tau_{j} = 0 \, ,
	\end{equation}
	and a vector field $v^{i}$ such that
	\begin{equation}
		v^{i}\tau_{i} = 1 \, .
	\end{equation}
	The condition in Eq.~\eqref{eq-cond-tau} implies that $\tau$ is hypersurface orthogonal. The volume element on $\M$ is given by
	\begin{equation}
	\widehat{\epsilon} = \tau \wedge \widetilde{\epsilon} \, , \label{NC-volume-element}
	\end{equation}
	where $ \widetilde{\epsilon}$ is the volume element on $\Sigma$.
	
	Note that we are not provided with a metric on $\M$. 
This means that the connection, and hence covariant derivative, on $\M$ can not be defined using the standard formulae from any textbook of GR, such as \cite{Wald}. To remedy this, we first define a unique tensor $h_{ij}$ by demanding 
	\begin{equation}
		h_{ij} v^{j} = 0 \, , \,\,\,\,\,\,\,\,\,\,\,\,\,\,\,\,\,\,\,\,\,\,\,\,\,\,\,\,\,\,\,\,\, h^{ik}h_{kj} = \delta^{i}_{\, j} -v^{i}\tau_{j} \equiv P^{i}_{\, j} \, . \label{eq-cond-h}
	\end{equation}
	Demanding `metric compatibility' conditions
	\begin{equation}
		D_{i}\tau_{j} = \, 0 \, , \,\,\,\,\,\,\,\,\,\,\,\,\,\,\,\,\,\,\,\,\,\,\,\,\,\,\,\,\,\,\,\,\, D_{i}h^{jk} = 0 \, ,
	\end{equation}
	and `curl-freeness' of $v^{i}$ 
	\begin{equation}
	D^{i}v^{j} - D^{j}v^{i} = \, 0 \, ,
	\end{equation}
	yields the unique connection \cite{son-14}
	\begin{equation}
		\nGamma^{i}_{\, jk} = v^{i}\partial_{j}\tau_{k} + \frac{1}{2}h^{il}\left( \partial_{j}h_{kl} + \partial_{k}h_{jl} - \partial_{l}h_{jk}\right) \, . \label{eq-nc-conn}
	\end{equation}
	Note that this connection has torsion unless $\tau_{i}$ is closed:
	\begin{align}
		T^{i}_{\, jk} \equiv& \, \nGamma^{i}_{\, jk} - \nGamma^{i}_{\, kj} \, ,\\
		=& \, v^{i}(d\tau)_{jk} \, . \label{eq-tor}
	\end{align}
	
	The Lie derivative of $h_{ij}$ along $v^{i}$
	\begin{align}
		B^{(v)}_{ij} \equiv& \frac{1}{2}\mathcal{L}_{v}h_{ij} \, , \label{eq-nc-extrinsic}
	\end{align}
	is transverse, that is $v^{i}B^{(v)}_{ij} = 0 \, .$
	
	\subsection{Global-time coordinates on $\M$}
	All the tensors and equations in the last section holds in any coordinate system on $\M$. However, we may choose a coordinate system , $x^{i} = \{t,x^{A}\}$, where $x^{A}$ are coordinates on $\Sigma$, and $x^{0} = t$ is the foliation parameter. These system of coordinates are called \emph{`global-time coordinates'} (GTC) in \cite{son-13,son-14}. In these coordinate systems, $\tau_{i}$ is of the form
	\begin{equation}
		\tau_{i} = N (dt)_{i} \, , \label{eq-nc-n-gtc}
	\end{equation}
	for some lapse function $N \ne 0$, where we have used the fact that $\tau$ is hypersurface orthogonal. The condition $h^{ij}\tau_{i} = 0$ means $h^{00} = h^{0A} = 0$, and $h^{AB} = \q^{AB}$, where $\q^{AB}$ is the inverse Riemannian metric on $\Sigma$. That is
	\begin{equation}
		h^{ij} = 
		\begin{pmatrix}
			0 & 0\\
			0 & \q^{AB}
		\end{pmatrix} \, . \label{eq-nc-h-inv-gtc}
	\end{equation}
	The condition $v^{i}\tau_{i} = 1$ means that
	\begin{equation}
		v^{i} = \begin{pmatrix}
			N^{-1} \\ N^{-1} s^{A} \end{pmatrix} \, . \label{eq-nc-v-gtc}
	\end{equation}
	Eq.~(\ref{eq-cond-h}) implies that $h_{ij}$ is of the form
	\begin{equation}
		h_{ij} = \begin{pmatrix}
			s^{2} & -s_{A} \\ -s_{B} & \q_{AB} \end{pmatrix} \, ,\label{eq-nc-h-gtc}
	\end{equation}
	where $s_{A} = \q_{AB}s^{B}$, and $s^{2} = s_{A}s^{A}$. The volume element in Eq.~\eqref{NC-volume-element} then becomes
	\begin{equation}
		\widehat{\epsilon} = dt \wedge d^{2}x \, N\sqrt{\q} \, .
	\end{equation}
	
	\subsection{Currents as response to the geometry}
	For a relativistic field theory, the energy momentum tensor is defined as the variation of the action with respect to the metric to which the theory is coupled. Similarly, the currents in a non-relativistic theory can be defined as the variation of the action with respect to the NC geometry \cite{son-13,son-14,Jensen:2014aia,Hartong:2014pma,Geracie:2015xfa}. Assume that we have an action as a functional of the NC data, $I_{NC}[\tau,h,v]$, and we want to compute its variation under a diffeomorphism generated by the vector field $\xi^{i}$. The variations of $\tau_{i}$, $v^{i}$, and $h^{ij}$ are
		\begin{align}
		\delta \tau_{i} = \mathcal{L}_{\xi}\tau_{i} =& \, \tau_{j}D_{i}\xi^{j} - T^{k}_{\, ij}\tau_{k}\xi^{j} \, ,\label{eq-var-tau} \\
		=& \, \tau_{j}D_{i}\xi^{j} - (d\tau)_{ij}\xi^{j} \, , \\
		\delta v^{i} = \mathcal{L}_{\xi}v^{i} =& \, \xi^{j} D_{j}v^{i} - v^{j} D_{j}\xi^{i}+T^{i}_{\,jk}v^{j}\xi^{k} \, , \\
		\delta h^{ij} = \mathcal{L}_{\xi}h^{ij} =& \, D^{i}\xi^{j} + D^{j}\xi^{i} + (T^{i\,\, j}_{\,\,k} + T^{j\,\, i}_{\,\, k})\xi^{k} \, .
		\end{align}
	Note that only those variations are allowed for which $\delta(v^{i}\tau_{i}) = 0$ and $\delta(h^{ij}\tau_{j})=0$. This means that we can keep $\delta \tau_{i}$ unconstrained, but demand that \cite{son-14} 
	\begin{align}
		\delta v^{i} =& - v^{i}v^{j}\delta \tau_{j} + \delta u^{i} \, ,\label{eq-var-u}\\
		\delta h^{ij} =& - v^{i}\delta \tau^{j} - v^{j}\delta \tau^{i} - \delta H^{ij} \, , \label{eq-var-H}
	\end{align}	
	where $\tau_{i}\delta u^{i} = 0$ and $\tau_{i} \delta H^{ij} = 0 \, .$ The energy current $\E^{i}$, the stress tensor $\Theta^{ij}$, and the momentum one-form $P_{i}$, are then defined in terms of the variation of the action, $I_{NC}[\tau,h,v]$,
	\begin{equation}
		\delta I_{NC} \equiv \int dt d^{2}x \, N\sqrt{\q} \, \left\{\frac{1}{2}\Theta^{ij}\delta H_{ij} - \E^{i}\delta \tau_{i} - P_{i}\delta u^{i} \right\} \, . \label{eq-var-act-NS-int}
	\end{equation}
	Note that the stress tensor and momentum one-form are transverse:
	\begin{equation}
		\Theta^{ij}\tau_{i} = 0 \, , \,\,\,\,\,\,\,\,\,\,\,\,\,\,\,\,\,\,\,\,\,\,\,\,\,\,\,\,\,\,\,\,\, P_{i}v^{i} = 0 \, . \label{eq-current-trans}
	\end{equation}
	Inserting the variations from Eqs.~\eqref{eq-var-tau}-\eqref{eq-var-H} in Eq.~(\ref{eq-var-act-NS-int}), and performing the integration by parts lead to \cite{son-14,Jensen:2014aia,Geracie:2015xfa} 
	\begin{equation}
		\delta I_{NC} = \int dt d^{2}x \, N\sqrt{\q} \, \, \xi^{j}  \left\{-D_{i}(P_{j}v^{i})-P_{i}D_{j}v^{i} - (D_{i}-T^{k}_{\, ki})\Theta_{j}^{\, i}+\tau_{j}(D_{i}-T^{k}_{\, ki})\E^{i} + (d\tau)_{ij}\E^{i}\right\} \, . \label{eq-var-act-NC-fin}
	\end{equation}
	This is an important result that we will use in Sec.~\ref{sec-screen-eqs} to write the covariant form of the screen equations, Eqs.~\eqref{screen-eq-scalar}-\eqref{screen-eq-vector}. We will also use the general form of the variation in Eq.~\eqref{eq-var-act-NS-int} to identify the currents on holographic screens. 
	
	\section{Newton-Cartan geometry on holographic screens} \label{sec-nc-on-screen}
	
	This section is the most important part of this paper. In this section, we will combine the ideas from the last two sections to propose that the geometry of the holographic screen, $H$, can be described by the NC geometry. In particular, we suggest that the Riemannian metric on the leaves, $\sigma(r)$, the vector field, $h^{i}$, normal to the leaves, and a non-vanishing scalar $\alpha$ in Eq.~\eqref{eq-lc-h}, play the role of NC data on $H$.
	
	The foliation structure of $H$ into leaves allows us to pick a coordinate system on $H$:  $\{r,x^{A}\}$, where $r$ is the foliation parameter, and $x^{A}$ are angular coordinates on $\sigma(r)$. These are the GTC of Sec.~(\ref{sec-nc-rev}). The foliation parameter, $r$, acts as the `\emph{screen time}'. One can always define a one-form, $(dr)_{i}$, which is normal to the leaves. We use this to write a one-form 
	\begin{equation}
	\tau_{i} =  \alpha (dr)_{i} \, . \label{eq-H-n}
	\end{equation}
	This means that $\alpha$ plays the role of the lapse function, $N$. This proposal would have been inconsistent if $\alpha$ were allowed to vanish anywhere. Recall that $\alpha$ is indeed non-vanishing,
	as was proven in \cite{bousso-engenhardt} assuming certain generic conditions.
	
	
		
	The second ingredient in the NC data is the vector field $v^{i}$, such that $v^{i}\tau_{i} = 1$. We use the vector field $h^{i}$ to define $v^{i} \equiv \alpha^{-1} h^{i}$. The coordinate representation of this vector field in the GTC, $\{r,x^{A}\}$, is
	\begin{equation}
		v^{i} = \begin{pmatrix} \alpha^{-1} \\ 0 \end{pmatrix} \, . \label{eq-H-v}
	\end{equation}
	Finally, we use the metric on the leaves, $\q_{AB}$, and its inverse $\q^{AB}$, to construct a $(2,0)$ symmteric tensor on $H$, whose coordinate representation is
	\begin{equation}
		h^{ij} = \begin{pmatrix} 0 & 0 \\ 0 & \q^{AB} \end{pmatrix} \, . \label{eq-H-met-inv}
	\end{equation}
	We combine $\tau_{i}$ in Eq.~(\ref{eq-H-n}), $v^{i}$ in Eq.~(\ref{eq-H-v}), and $h^{ij}$ in Eq.~(\ref{eq-H-met-inv}) to put a NC structure on the screen. This allows us to define geometric quantities like connection and covariant derivatives on $H$. The non-zero components of the connection in Eq.~(\ref{eq-nc-conn}) are
	\begin{align}
		\nGamma^{0}_{\, i 0} =& \, \partial_{i}\log\alpha \, ,\\
		\nGamma^{A}_{\, B 0} =& \, \nGamma^{A}_{\, 0 B} = \, \frac{1}{2} \q^{AC}\partial_{0}\q_{BC} \, ,\\
		\nGamma^{A}_{\, BC} =& \, \widetilde{\Gamma}^{A}_{\, BC} \equiv \frac{1}{2} \q^{AD}\left( \partial_{B}\q_{CD} + \partial_{C}\q_{BD} - \partial_{D}\q_{BC}\right) \, ,
	\end{align}
	where $\widetilde{\Gamma}^{A}_{\, BC}$ are the Christoffel symbols of the metric $\q_{AB}$. 
	
%
%
%
%
	Note that we have only presented the NC data in the GTC. The representation of $\tau_{i}$, $v^{i}$, and $h^{ij}$ in any arbitrary coordinate system can be determined using the transformation law of tensors. This is an important result, as this allows us to describe the geometry of $H$ without using the foliation structure of $H$.
	
	Also note that the scalar $\beta$ in Eq.~\eqref{eq-lc-h}, unlike $\alpha$, does not appear in Eqs.~\eqref{eq-H-n}-\eqref{eq-H-met-inv}. This means that the intrinsic geometry of $H$ is not dependent on $\beta$. This is consistent with the interpretation of $\beta$ as the ``slope'' of $H$ \cite{bousso-moosa}. That is, it tells us how much the screen bends in the spacetime it is embedded in. Therefore, $\beta$ is an extrinsic quantity.
	
%
	To finish this section, we want to compare the NC geometry with the geometry described by the induced metric for screens which are nowhere null. The advantage of NC geometry is that the foliation structure of screen is implicit in it through one-form field $\tau$. Another advantage of using NC geometry will be apparent in Sec. (\ref{sec-var-calc}), where we will find that the boundary conditions on NC data are more suitable for holographic screens than the boundary conditions on the induced metric.
	
	\subsection{Covariant screen equations} \label{sec-screen-eqs}
	
	The evolution equations of holographic screen, Eqs.~\eqref{screen-eq-scalar}-\eqref{screen-eq-vector}, are the decomposition of constraint equations, Eq.~\eqref{eq-einstein}, into directions normal and transverse to the leaves. These equations are invariant under the foilation-preserving diffeomorphisms. In this section, we will show that we can use the NC geometry to write a covariant version of these equations. 
	
	Before we derive the covariant screen equations, recall how the constraint equations, Eq.~\eqref{rel-open-sys}, for the hypersurface of a fixed signature arise from diffeomorphism invariance. Imagine taking the spacetime and truncating it at the hypersurface. As a result, the Einstein-Hilbert (EH) action has to be compensated with the Gibbons-Hawking-York (GHY) boundary term living on the hypersurface,
	\begin{align}
	I =& \, I_{EH}[g] + I_{GHY}[g,\gamma] \, .
	\end{align}
	The variation of this action is~\cite{brown-york} 
	\begin{align}
		\delta I =& \, \frac{1}{16\pi}\int_{M} d^{4}x \, \sqrt{g} \, G_{ab}\delta g^{ab} + \frac{1}{16\pi}\int_{\partial M} d^{3}x \, \sqrt{\gamma} \, \left( \widehat{K}_{ij} - \widehat{K} \, \gamma_{ij} \right) \, \delta \gamma^{ij} \, .
	\end{align}
Now we take a vector field tangent to the hypersurface, $\xi^{i}$, and generate diffeomorphism by it. The variation of the total action under this transformation is  
	\begin{equation}
		\delta I =  -\frac{1}{8\pi}\int_{M} d^{4}x \, \sqrt{g} \, G_{ab}\nabla^{a}\xi^{b} - \frac{1}{8\pi} \int_{\partial M} d^{3}x \, \sqrt{\gamma} \, \left( \widehat{K}_{ij} - \widehat{K} \, \gamma_{ij} \right) \, \widehat{D}^{i}\xi^{j} \, ,
	\end{equation}
	where $\xi^{a}= e^{a}_{i}\xi^{i}$ is the push-back of the vector field on the full spacetime. Performing the integration by parts yields
	\begin{equation}
		\delta I = \frac{1}{8\pi}\int_{M} d^{4}x \, \sqrt{g} \, \xi^{b} \, \nabla_{a}G^{ab}   -\frac{1}{8\pi}\int_{\partial M} d^{3}x \, \sqrt{\gamma} \, \xi^{i} \, \Big[G_{ab}n^{a}e^{b}_{i} - \widehat{D}_{j} \left( \widehat{K}_{ij} - \widehat{K} \, \gamma_{ij} \right) \Big] \, .
	\end{equation}
	Demanding that the action is invariant under diffeomorphisms generated by $\xi^{i}$ gives us Bianchi identity and Gauss-Codazzi constraint. The latter when combined with the Einstein equations becomes Eq.~\eqref{rel-open-sys}.
	
	Now we repeat the similar analysis for hologrpahic screens, $H$. That is, we take a spacetime in which the holographic screen is embedded in, and truncate it at $H$. The variational principle requires that we add a boundary action on $H$, whose variation cancels the boundary term in the variation of the EH action. In the next section, we will follow \cite{booth-fairhurst} and study this variational problem. For now, 	
%
%
simply assume that there exists a screen action as a functional of NC data such that the combination
	\begin{equation}
		I = I_{EH}[g] -  I_{H}[g,\tau,h,v] \, .
	\end{equation}
	has a well-defined variation, and is given by 
	\begin{equation}
		\delta I = \frac{1}{16\pi}\int_{M} d^{4}x \, \sqrt{g} \, G_{ab}\delta g^{ab} -  \int_{H} dr d^{2}x \, \alpha \, \sqrt{\q} \, \left\{\frac{1}{2}\Theta^{ij}\delta H_{ij} - \E^{i}\delta \tau_{i} - P_{i}\delta u^{i} \right\}\, , \label{eq-var-total-gravity-action-with-screen}
	\end{equation}
	where we have used Eq.~\eqref{eq-var-act-NS-int}.  If we take the variation to be a diffeomorphism generated by a vector field $\xi^{i}$, on $H$, then we get
	\begin{align}
		\delta I =& \,  -\frac{1}{8\pi}\int_{M} d^{4}x \, \sqrt{g} \, G_{ab}\nabla^{a}\xi^{b} \nonumber\\ & -  \int_{H} dr d^{2}x \, \alpha\sqrt{\q} \, \, \xi^{j}  \left\{-D_{i}(P_{j}v^{i})-P_{i}D_{j}v^{i} - (D_{i}-T^{k}_{\, ki})\Theta_{j}^{\, i}+\tau_{j}(D_{i}-T^{k}_{\, ki})\E^{i} + (d\tau)_{ij}\E^{i}\right\} \, ,
	\end{align}
	where we have used Eq.~\eqref{eq-var-act-NC-fin} in the second term. 
	Performing the integration by parts, and demanding $\delta I = 0$ leads to a Bianchi identity, and a constraint equation which when combined with Einstein equations becomes
	\begin{equation}
		D_{i}(P_{j}v^{i}) + P_{i}D_{j}v^{i} + (D_{i}-T^{k}_{\, ki})\Theta_{j}^{\, i} - \tau_{j}(D_{i}-T^{k}_{\, ki})\E^{i} - (d\tau)_{ij}\E^{i} =  \alpha^{-1} \, T_{ab}n^{a}e^{b}_{j} \, . \label{eq-screen-conv}
	\end{equation}
	In the following, we will show that this tensor equation is a covariant version of the screen equations, Eqs.~\eqref{screen-eq-scalar}-\eqref{screen-eq-vector}. In particular, we will show that this equation reduces to Eqs.~\eqref{screen-eq-scalar}-\eqref{screen-eq-vector} when we expand it in the GTC. Note that Eq.~\eqref{eq-current-trans} implies that the momentum one-form $P_{i}$, stress tensor $\Theta^{ij}$,  and energy current $\E^{i}$ have the following representation in the GTC
	\begin{align}
	P_{i} =&\, \big(0, P_{A} \big) \, , \label{eq-ans-p}\\
	\Theta^{ij} =&\, \begin{pmatrix} 0 & 0\\ 0 & \Theta^{AB} \end{pmatrix} \, , \label{eq-ans-t}\\
	\E^{i} =&\, \begin{pmatrix} \E^{0} \\ \E^{A} \end{pmatrix} \, .\label{eq-ans-e}
	\end{align}

	The contraction of Eq.~(\ref{eq-screen-conv}) with $v^{i}$ is
	\begin{align}
		\alpha^{-1} \, T_{ab}n^{a}v^{b} =& \, - (D_{i}-T^{k}_{\, ki})\E^{i} + T^{k}_{\, ki}\E^{i} - \Theta_{i}^{\, j}D_{j}v^{i} \, \\
		=& \, -D_{i}\E^{i} + 2 T^{k}_{\, ki}\E^{i} -\Theta^{ij}B^{(v)}_{ij} \,. \label{eq-screen-v}
	\end{align}
	%
	By writing this equation in the GTC, and using the relation $v^{i} = \alpha^{-1} h^{i}$, we get
	\begin{equation}
		T_{ab}n^{a}h^{b} = -\alpha \mathcal{L}_{h}(\alpha \E^{0}) - \alpha^{2} \E^{0}\theta^{(h)} - \widetilde{D}_{A}(\alpha^{2} \E^{A}) - \alpha \, \Theta^{AB}B^{(h)}_{AB} \, . \label{eq-sc-scalar} 
	\end{equation}
	Similarly, the contraction of Eq.~(\ref{eq-screen-conv}) with $h^{ij}$ is
	\begin{align}
		\alpha^{-1} \, T_{ab}n^{a}e^{b}_{j} h^{ij} =& \, h^{ij} \, \mathcal{L}_{v}P_{j} + h^{ij} P_{j} \theta^{(v)} + (D_{j}-T^{k}_{\, kj})\Theta^{ij} - h^{ij}(d\tau)_{kj}\E^{k} \, . 
	\end{align}
	Writing this equation in the GTC leads to 
	\begin{equation}
		T_{ab}n^{a}e^{b}_{A} = \q_{A}^{\, B}\mathcal{L}_{h}p_{B} + p_{A}\theta^{(h)} + \widetilde{D}_{B}(\alpha \Theta_{A}^{\, B}) + \alpha \E^{0}\widetilde{D}_{A}\alpha \, . \label{eq-sc-vector}
	\end{equation}
	With the identifications
	\begin{align}
		P_{A} =& \, \frac{1}{8\pi} \, \Omega_{A} \, , \label{eq-iden-p}\\
		\E^{0} =& \, -\frac{1}{8\pi \, \alpha}\theta^{(l)} \, = - \frac{1}{8\pi \, \alpha^{2}} \theta^{(h)} , \label{eq-iden-en}\\
		\E^{A} =& \, -\frac{1}{8\pi \, \alpha^{2}}\widehat{\Omega}^{A} \, , \label{eq-iden-c}\\
		\Theta_{AB} =& \, \frac{1}{8\pi \, \alpha} \left(B^{(n)}_{AB} - \theta^{(n)}\q_{AB} - \widetilde{\kappa}\q_{AB}\right) \, . \label{eq-iden-s}
	\end{align}
	Eq.~(\ref{eq-sc-scalar}) and Eq.~(\ref{eq-sc-vector}) reduces to Eq.~\eqref{screen-eq-scalar} and Eq.~\eqref{screen-eq-vector} respectively. This confirms our claim that  Eq.~\eqref{eq-screen-conv} is a covariant generalization of the screen evolution equations. We view this as an application of the Newton-Cartan formalism to describe the geometry of the screen. However at this point, the identifications that we have made in Eqs.~\eqref{eq-iden-p}-\eqref{eq-iden-s} seem arbitrary. In the next section, we will follow \cite{booth-fairhurst} to write the variation of the total action in a form similar to Eq.~\eqref{eq-var-total-gravity-action-with-screen}. This will allow us to verify the identifications that we have made for $P_{A}$, $\E^{0}$, and $\Theta_{AB}$. 
	
	
	
		\section{Derivation of $P_{A}$, $\Theta_{AB}$, and $\E^{0}$ from gravitational action} \label{sec-var-calc}
	
	In the spacetime with boundaries, the Einstein-Hilbert action has to be compensated with a boundary term which depends on the boundary conditions. For instance, Dirichlet conditions on a boundary of fixed signatures leads to Gibbons-Hawking-York term. Brown and York studied this problem in \cite{brown-york}, where they assumed that the spacetime is bounded by four boundaries: an outer timelike, an inner timelike, a future spacelike, and a past spacelike boundaries. This analysis was generalized in \cite{PhysRevD.47.3275}, where the corner contributions from the interaction of two different boundaries were also considered. This was further generalized in \cite{booth-fairhurst}, where the inner timelike boundary was replaced with the codimension-$1$ hypersurface
	which
	\begin{itemize}
		\item is foliated by the codimenion-$2$ marginally trapped or anti-trapped closed surfaces, 
		\item is of indefinite signature. 
	\end{itemize}
	Note that these are exactly the properties of holographic screens. In this section, we will follow the calculations of \cite{booth-fairhurst}. However, our analysis will be different from that of \cite{booth-fairhurst} in two ways. As we only care about the boundary contributions to the Einstein-Hilbert action coming from the holographic screen, we will assume that there are no other boundaries of the spacetime. The other, and more important, difference in the two analysis is that we will be assuming slightly different boundary conditions. 
	
	We now discuss the boundary conditions that we are imposing on the holographic screens. The Dirichlet condition on the induced metric is, of course, not suitable as the signature of  screens is allowed to change. There is another reason for why Dirichlet conditions are inapplicable. Note that we are only allowed to impose six boundary conditions on the screen. If we fix the induced metric on the screen, then we can not fix any other geometric data. However, holographic screens are special as the leaves have vanishing null-expansion, $\theta^{(k)} = 0$. Following \cite{booth-fairhurst}, we do not impose any boundary conditions on the induced metric. Rather, we assume a Dirichlet condition on  $\theta^{(k)}$. 
	This boundary condition also fixes the foliation structure of the holographic screen. 
%
%
	
	We still need to choose five more boundary conditions. We fix three of those by imposing Dirichlet condition on the metric on the leaves, $\q_{AB}$, which also fixes its inverse, $\q^{AB}$. So far, our four boundary conditions are same as those of \cite{booth-fairhurst}. Next, we choose to impose boundary conditions on the variation of the evolution vector field, $h^{i}$. Note that the normalization $h^{i}(dr)_{i} = 1$ implies that $\delta h^{i}$ is transverse,
	\begin{equation}
	\delta h^{i} = \, e^{i}_{A} \, \delta s^{A} \, .
	\end{equation}
	We now impose the Dirichlet conditions on $s^{A}$. These are our last two boundary conditions. Note that these conditions are different from those of \cite{booth-fairhurst} where they had imposed Dirichlet conditions on $\Omega_{A}$. As we will see below, our boundary conditions not only result in the well-defined variation of the action, it will also help us to compare the variation of the total action with Eq.~\eqref{eq-var-total-gravity-action-with-screen}.

	As emphasized in \cite{booth-fairhurst}, we also have to fix the `length' of the null vector $k^{a}$. Note that this is not an additional boundary condition. We need to impose this condition because the extrinsic quantities like $\widetilde{\kappa}$ and $\Omega_{A}$ depend on the `length' of the null vectors \cite{bousso-moosa}. This is equivalent to the case of the boundary with a fixed signature, where we fix the norm of the normal vector to be $\pm 1$. Without fixing the norm of the normal vector, the extrinsic curvature and hence the GHY boundary term have ambiguities, and are not well defined. We can fix the `length' of the null vectors by imposing the Dirichlet condition on $\alpha \, .$

	After discussing the boundary conditions in detail, we now consider the variation of EH action on the spacetime with the holographic screen as the boundary. We copy from \cite{booth-fairhurst}
	\begin{align}
		\delta I_{EH}[g] =& \, \frac{1}{16\pi} \int_{M} d^{4}x \,\sqrt{g} \, G_{ab} \, \delta g^{ab} \nonumber\\ &+ \frac{1}{16\pi} \int_{H} dr d^{2}x \, \sqrt{\q} \, \left( -2 (\delta \theta^{(n)}) - (\delta\q_{AB})B^{(n)AB} - 2 (\delta\alpha) \theta^{(l)} - 2 (\delta\widetilde{\kappa}) + 2(\delta h^{i}) \omega_{i} \right) \, ,
	\end{align}
	which we re-write as \cite{booth-fairhurst}
	\begin{align}
		\delta I_{EH}[g] =& \, \frac{1}{16\pi} \int_{M} d^{4}x \,\sqrt{g} \, G_{ab} \, \delta g^{ab} -\frac{1}{8\pi} \delta \left\{ \int_{H} dr d^{2}x \, \sqrt{\q} \, \Big(\widetilde{\kappa}+\theta^{(n)}\Big)\right\} \nonumber\\	 &- \frac{1}{8\pi} \int_{H} dr d^{2}x \, \sqrt{\q} \, \left( \frac{1}{2} (\delta\q_{AB})\Big(B^{(n)AB}-\theta^{(n)}\q^{AB}-\widetilde{\kappa}\q^{AB}\Big) - (\delta s^{A})\Omega_{A} + (\delta\alpha)\theta^{(l)} \right) \, . \label{var-bf-two}
	\end{align}
	Note that if we impose our boundary conditions, then the second line in the above equation vanishes. The second term in the first line is a total variation, and should be added to EH action as the boundary term. Therefore, the total gravitational action in the spacetime with holographic screen as the boundary is
	\begin{equation}
		I = \frac{1}{16\pi} \, \int_{M} d^{4}x \, \sqrt{g} \, R + \frac{1}{8\pi} \, \int_{H} dr d^{2}x \, \sqrt{\q} \, (\widetilde{\kappa}+\theta^{(n)}) \, .
	\end{equation}
	The variation of this total action follows from Eq.~\eqref{var-bf-two}
	\begin{align}
		\delta I =& \, \frac{1}{16\pi} \int_{M} d^{4}x \,\sqrt{g} \, G_{ab} \, \delta g^{ab} \nonumber\\ & \, - \frac{1}{8\pi} \int_{H} dr d^{2}x \, \alpha \, \sqrt{\q} \, \left( \frac{1}{2} (\delta\q_{AB})\frac{1}{\alpha}\Big(B^{(n)AB}-\theta^{(n)}\q^{AB}-\widetilde{\kappa}\q^{AB}\Big) - \frac{1}{\alpha}(\delta s^{A})\Omega_{A} + (\delta\alpha)\frac{1}{\alpha}\theta^{(l)} \right) \, . \label{eq-var-BF}
	\end{align}

	By comparing the above expression for the total variation with Eq.~\eqref{eq-var-total-gravity-action-with-screen} (or Eq.~\eqref{eq-var-act-NS-int}), we can deduce the expressions for momentum one-form $P_{i}$, stress tensor $\Theta^{ij}$, and energy current $\E^{i}$. However, all the calculation that we have done in this section are in the GTC gauge, Eq.~\eqref{eq-nc-n-gtc}-\eqref{eq-nc-h-gtc}. In this gauge, the variation $\delta\tau_{i}$, $\delta u^{i}$ in Eq.~\eqref{eq-var-u}, $\delta H^{ij}$ in Eq.~\eqref{eq-var-H}, are of the form
	\begin{align}
	\delta \tau_{0} =& \, \delta \alpha \, ,\\
	\delta \tau_{A} =& \, 0 \, ,\\
	\delta u^{i} =& \, \alpha^{-1} \,e_{A}^{i} \delta s^{A} \, \\
	\delta H^{ij} =& \, e_{A}^{i} e_{B}^{i} \, \delta \q^{AB} \, .
	\end{align}
	With these variations, Eq.~\eqref{eq-var-total-gravity-action-with-screen} becomes
		\begin{equation}
		\delta I = \frac{1}{16\pi}\int_{M} d^{4}x \, \sqrt{g} \, G_{ab}\delta g^{ab} -  \int_{H} dr d^{2}x \, \alpha \, \sqrt{\q} \, \left\{\frac{1}{2}\Theta^{AB}\delta \q_{AB} - \alpha^{-1} P_{A} \, \delta s^{A} - \E^{0} \, \delta\alpha \right\}\, . \label{eq-var-total-gravity-action-with-screen-GTC}
		\end{equation}
	Comparison of this expression with Eq.~\eqref{eq-var-BF} leads us to conclude that
	\begin{align}
		P_{A} =& \, \frac{1}{8\pi} \, \Omega_{A} \, ,\\
		\E^{0} =& \, -\frac{1}{8\pi \, \alpha}\theta^{(l)} \, = - \frac{1}{8\pi \, \alpha^{2}} \theta^{(h)} ,\\ 
		\Theta_{AB} =& \, \frac{1}{8\pi \, \alpha} \left(B^{(n)}_{AB} - \theta^{(n)}\q_{AB} - \widetilde{\kappa}\q_{AB}\right) \, , 
	\end{align}
	which agrees with the identifications that we made in Eqs.~\eqref{eq-iden-p}-\eqref{eq-iden-en} and in Eq.~\eqref{eq-iden-s}. However, note that we have failed to derive the expression for $\E^{A}$ using the above formalism. The reason behind this failure is that we have been working in the GTC gauge, in which $\delta\tau_{A} = 0$. To derive the spatial components of the energy current, $\E^{A}$, we need to vary the action of the theory with respect to $\tau_{A}$. This requires going away from the GTC at least to the first order. It is not clear how to perform the above analysis in a fully covariant way. We leave understanding this task to future work. 
	

	\section{Discussion}
	
	We have seen that the induced metric on a holographic screen is degenerate, and is not applicable to define geometric structure such as connection and covariant derivative. We have proposed that the intrinsic geometry should, instead, be described by the Newton-Cartan geometry. We have used some results from \cite{bousso-engenhardt} to argue that this proposal is consistent. Using the NC connection on $H$, we have presented the screen evolution equations of \cite{bousso-moosa} into a covariant form. There are many possible directions which we intend to pursue in the future. We briefly discuss these in the following.
	
	The natural next step would be to couple a (non-relativistic) field theory to holographic screen. The coupling of field theories to Newton-Cartan geometry has been studied in \cite{Jensen:2014aia,Hartong:2014pma,Geracie:2015xfa}. It would be interesting to know if there are any constraints on the field theory that we can define on the screen. One possible way to find these constraints might be to compare the entanglement structure of the field theory with the screen entanglement conjecture of \cite{weinberg-sanches}.
		 
	In AdS spacetime, metric near the conformal boundary can be written as a power series in the bulk coordinate \cite{Graham:1999pm}. This type of expansion has also been studied in \cite{Christensen:2013lma} for Lifshitz gravity, where the geometry of the boundary is Newton-Cartan. It would be fascinating if a metric of a spacetime can be expanded in a similar fashion near the screen. One-to-one correspondence between the screen and the null foliation of the spacetime suggests that the affine parameter on null slices can be used as a bulk coordinate.
	
	As discussed in Sec.~(\ref{sec-var-calc}), we also plan to understand how to generalize the calculations of \cite{booth-fairhurst} without using the GTC gauge, Eq.~\eqref{eq-nc-n-gtc}-\eqref{eq-nc-h-gtc}. This would be important as this will allow us to derive the spatial components of the energy currents, $\E^{A}$, which we failed to do in Sec.~(\ref{sec-var-calc}).
	

%
	

\section*{Acknowledgements}

We would like to thank R.~Bousso, Z.~Fisher, S.~Leichenauer, N.~Obers, M.~Rangamani, A.~Shahbazi, and Z.~Yan for helpful discussions, and to Z.~Fisher and S.~Shaukat for useful feedbacks on a draft of this manuscript. This work was supported in part by the Berkeley Center for Theoretical Physics, by the National Science Foundation (award numbers 1521446 and 1316783), by FQXi, and by the US Department of Energy under Contract DE-AC02-05CH11231.


\bibliographystyle{utcaps}
\bibliography{refs}

\end{document}